\documentclass[final]{IEEEtran}

\usepackage{amsthm,amsmath,amssymb,mathtools,graphicx,multirow,amsmath,color,amsfonts}
\usepackage[noadjust]{cite}
\usepackage[update,prepend]{epstopdf}
\usepackage{subcaption}

\newcommand\numberthis{\addtocounter{equation}{1}\tag{\theequation}}

\usepackage{setspace}	
\allowdisplaybreaks 

\def\nbo{{\mathbf{o}}}

\def\nbr{{\mathbf{r}}}

\def\nbu{{\mathbf{u}}}

\def\nbx{{\mathbf{x}}}
\def\nby{{\mathbf{y}}}

\def\nb0{{\mathbf{0}}}
\def\nb1{{\mathbf{1}}}

\def\ncalI{{\mathcal{I}}}

\def\ncalM{{\mathcal{M}}}

\def\nbbE{{\mathbb{E}}}

\def\nbbP{{\mathbb{P}}}

\newtheorem{lemma}{Lemma}

\newtheorem{prop}{Proposition}

\newtheorem{remark}{Remark}

\def\R{\mathbb{R}}

\def\sinr{\mathtt{SINR}}			

\begin{document}
\title{Modeling of Dense CSMA Networks using Random Sequential Adsorption Process}

\author{
Priyabrata Parida and Harpreet S. Dhillon.  \vspace{-1cm}
\thanks{The authors are with Wireless@VT, Department of ECE, Virginia Tech, Blacksburg, VA. Email: \{pparida, hdhillon\}@vt.edu. The support of the US NSF (Grant ECCS-1731711) is gratefully acknowledged. \hfill}
}

\maketitle

\begin{abstract}
We model a dense wireless local area network (WLAN) where the access points (APs) employ carrier sense multiple access (CSMA)-type medium access control (MAC) protocol. 
In our model, the spatial locations of the set of {\em active} APs are modeled using the random sequential adsorption (RSA) process, which is more accurate in terms of the density of active APs compared to the Mat{\'e}rn hard-core point process of type-II (MHPP-II) commonly used for modeling CSMA networks.
Leveraging the theory of the RSA process from the statistical physics literature, we provide an approximate but accurate analytical result for the medium access probability $({\tt MAP})$ of the typical AP in the network. 
Further, we present a numerical approach to determine the pair correlation function $({\tt PCF})$, which is useful for accurate estimation of the interference statistics. Using the ${\tt PCF}$ result, we derive the signal-to-interference-plus-noise ratio $(\sinr)$ coverage probability of the typical link in the network.
We validate the accuracy of the theoretical results through extensive Monte Carlo simulations.
\end{abstract}
\vspace{-0.4cm}

\section{Introduction}
Owing to their ubiquity, WLAN or Wi-Fi networks play a pivotal role in meeting the ever-increasing global data demands.
Since these networks operate with minimal central coordination, MAC protocols that rely on local information are preferred.
One such MAC protocol is the CSMA that enables efficient spatial sharing of the frequency bands among the WLAN APs. Hence, it has become the {\it de facto} MAC protocol in the popular IEEE 802.11 standards. 
The capacity of such a network not only depends on the probability of an AP accessing the medium but also the network interference generated by all active APs.
Since these quantities strongly depend on the active AP locations in the network (which are an outcome of complex spatio-temporal interactions across all APs),  
one approach to analyzing such a network is to model these AP locations using an accurate point process. Later leveraging the known statistical properties of the process, different system metrics can be reliably estimated.
Indeed, this philosophy has gained significant traction in recent years to study a variety of wireless networks using various point processes from the stochastic geometry literature~\cite{Haenggi2013,andrews2016primer,dhillon2020poisson}.
However, in the case of the CSMA network, the problem is more complicated as the spatial inhibition among the active APs is modeled using hard-core processes, which are notoriously challenging to analyze.  

A popular hard-core process that has been used to model the CSMA network is the MHPP-II~\cite{nguyen2007, alfano2012new, wang2016, li2016modeling}. Although the derivation of exact results remains intractable using MHPP-II, the knowledge of its second-order statistics in closed form provides a reasonable degree of tractability to obtain approximate results.
However, it is well understood that MHPP-II underestimates the density of simultaneously active transmitters that subsequently leads to an underestimation of network interference~\cite{busson2009}.
To overcome this limitation of the MHPP-II, the RSA process has been proposed to model the CSMA network~\cite{busson2009, nguyen2012}. In~\cite{busson2009}, the authors demonstrate the accuracy of the RSA-based model over MHPP-II through a simulation-based study. In~\cite{nguyen2012}, the authors present the generating functional of the RSA process as a solution to a differential equation. Since the exact solution of the equation is numerically demanding, the authors present a few bounds on network performance metrics that are rather loose. 
In contrast to the limited works in the communications literature, the RSA process has received significant attention from the statistical physics community (cf.~\cite{talbot2000car} and the references therein) where this process is used to model real-world phenomenon such as deposition of colloidal particles on the surface of a substrate.

In this work, we revisit the problem of modeling a dense CSMA network using the RSA process. From the statistical physics literature, we recover a few useful results corresponding to the first and second-order statistics of the process and apply them, {\it albeit} with suitable modifications, to analyze the CSMA network.  Since the exact expressions for different metrics are intractable, we derive approximate but accurate theoretical results, which are amenable to faster numerical evaluations. Our main contributions are as follows: 

{\bf 1.} Owing to the contention-based medium access in a CSMA network,  the knowledge of the ${\tt MAP}$ of the typical AP is critical for network capacity estimation. Therefore, we extend the circular void probability result of the RSA process from the statistical physics literature to present an accurate theoretical expression for the ${\tt MAP}$ of the typical AP.
We also present the result for the density of the active APs in the network.

{\bf 2.} For statistical characterization of network interference, we resort to accurate approximation using the two-particle density function of the RSA process. Inspired by the results from the physics literature, we present a useful numerical approach to obtain the ${\tt PCF}$ of the RSA process. To facilitate tractable analysis, we also provide a parametric closed-form expression for the ${\tt PCF}$, where the desired parameters can be obtained through curve-fitting with respect to either the numerical or simulated ${\tt PCF}$. Using the ${\tt PCF}$ result, we approximate the conditional RSA process as a non-homogeneous PPP. This approximation helps us determine the $\sinr$ coverage probability.



\vspace{-0.25cm}

\section{System Model}\label{sec:SysMod}

{\bf \em  A. Network model:}
We consider the downlink (DL) of a  WLAN network where the locations of APs follow a homogeneous Poisson point process (PPP) $\Phi_A = \{\nbx_1, \nbx_2, \ldots\}$ with density $\lambda_a$. Since deployment of WLAN APs are unplanned, a PPP model is quite reasonable in this case.
{Further, we assume that the set of users served by each AP is randomly and uniformly distributed within a circle of radius $r_{\tt inh}$ centered at the corresponding AP. This assumption is motivated by the facts that WLAN networks are usually closed-access systems and the users are in the general vicinity of serving APs. 
Further, this assumption is a generalization of the {\em bipolar model} used for modeling serving distance in ad hoc networks.}
Our focus is on a WLAN system with a single channel. Further without loss of generality, the analysis is for a representative resource block with flat frequency response.
On the representative resource block, considering a fixed AP transmit power $P_t$, the received power at $\nby \in \R^2$ from an AP at $\nbx_i$ is 
\begin{align*}
P_r(\nby, \nbx_i) = P_t h(\nby, \nbx_i)l(\|\nby - \nbx_i\|),
\end{align*}
where $h(\nby, \nbx_i)$ is the multi-path gain of the link. Assuming that multi-path fading follows Rayleigh distribution, we model the gain as an independent and identically distributed (i.i.d.) exponential random variable with unit mean for each link.
Further, $l(\|\nby - \nbx_i\|)$ is the distance dependent path-loss between $\nby$ and $\nbx_i$. For simplicity, we assume that $l(\|\nby - \nbx_i\|) = \|\nby - \nbx_i\|^{-\alpha}$, where $\alpha > 2$ is the path-loss exponent.

\begin{figure}[!htb]
\vspace{-0.5cm}
  \centering
  \includegraphics[width=\columnwidth]{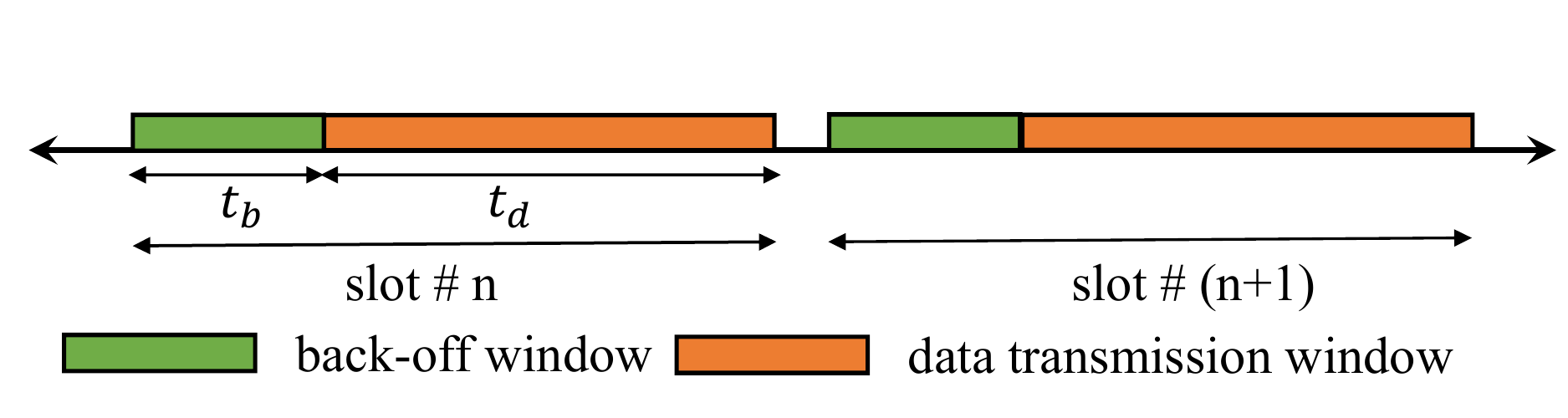}
  \caption{\footnotesize Example of two consecutive slots for the slotted CSMA considered in this work. The slots are represented from the perspective of the AP, i.e., the transmitter. During the gap between two consecutive slots, control signaling such as acknowledgement of correct reception from the user takes place.}
  \label{fig:slotted_csma}
\end{figure}

{\bf \em B. Contention-based medium access:}
As shown in Fig.~\ref{fig:slotted_csma}, we consider a slotted CSMA-type MAC protocol. The duration of each slot is $t_b + t_d$ time units, where the maximum back-off period is $t_b$ and the minimum data transmission duration is $t_d$.
At the beginning of a slot, each AP enters a back-off period that is uniformly distributed in $(0, t_b]$ and independent for each AP as well as the history of the back-off times in the previous slots.
In case the back-off times of two APs are the same, advanced protocols can be used to avoid  packet collision.
For our construction of the back-off procedure, the probability of such events is zero. Hence, we do not consider such events in our analyses.
Further, since there is no packet collision, we also do not consider exponential back-off in our analyses. 
We consider saturated traffic for each AP. Hence, during each slot, all the APs in the network participate in channel contention.
In this period, an AP continuously senses the channel to register the activity of nearby APs that may have smaller back-off periods and may have started data transmission.
If the AP infers the channel to be active by the time the back-off period ends, it waits for the next slot and then follows the same process.
In contrast, if the AP does not detect the presence of other APs in its {\em contention/inhibition region}, it transmits data until the end of the slot. The same process repeats at the beginning of each slot.
Note that during a slot, the duration of the data transmission period for the typical AP in the network is uniformly distributed in $[t_d, t_d + t_b)$. This is a consequence of the random back-off period and having a fixed slot duration. 

\begin{figure}[!htb]
  \centering
  \includegraphics[width=0.6\columnwidth]{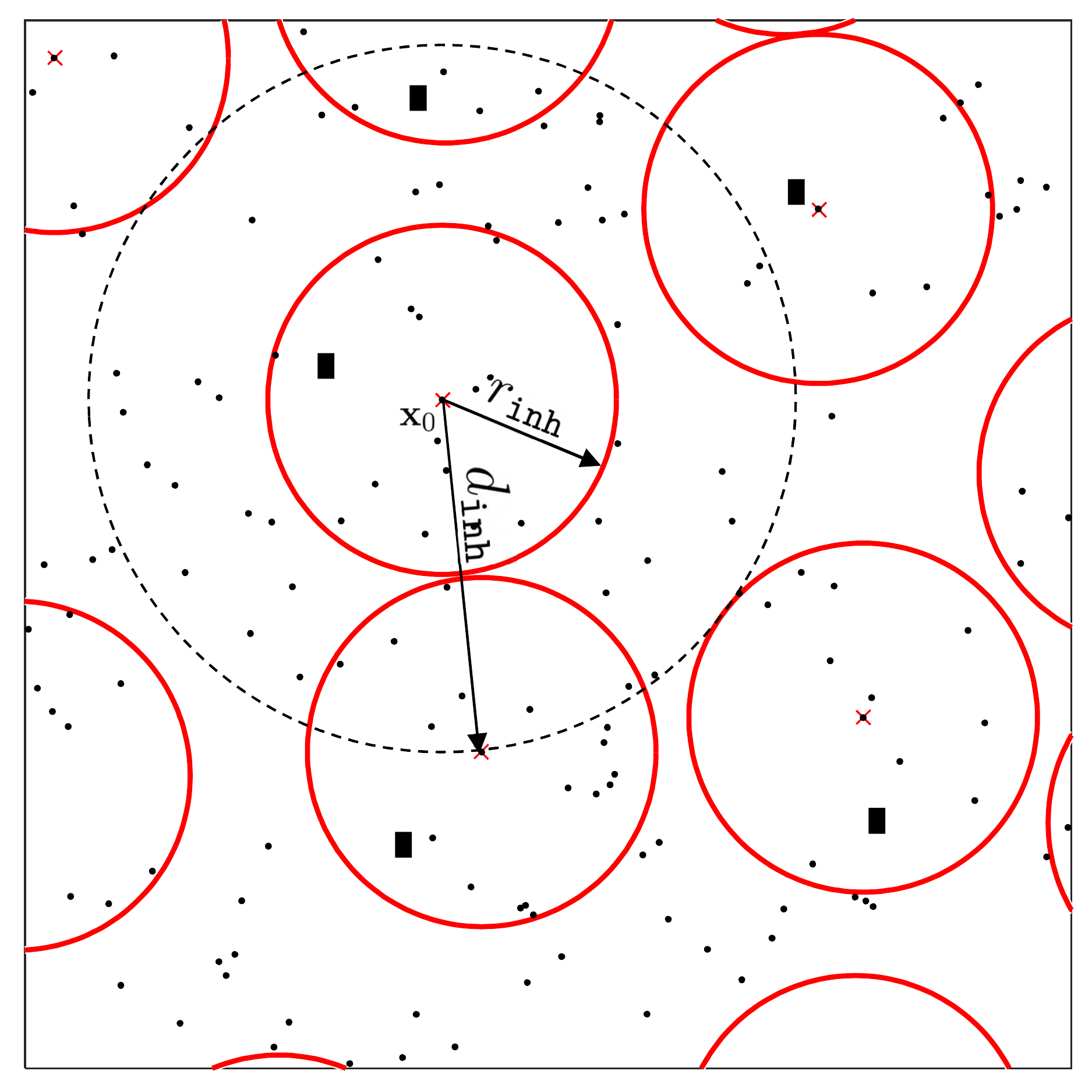}
  \caption{\footnotesize A representative illustration of the system. Each red circle is centered at an active AP illustrated by cross mark. APs that did not win contention are denoted by black-dots. A few representative user locations are illustrated by dark rectangles. The inhibition region of the typical AP at $\nbx_0$ is $B_{d_{\tt inh}} (\nbx_0)$ that is illustrated by the dotted black circle of radius $d_{\tt inh}$. For the typical AP, the serving users are uniformly distributed in $B_{r_{\tt inh}} (\nbx_0)$.}
  \label{fig:system_illus}
\end{figure}

During the back-off period, the typical AP can detect the presence of other active APs in its contention region through energy detection and/or preamble detection.
In this work, we only consider preamble detection. 
To successfully decode the preamble, we assume that the averaged received signal strength needs to be above the sensing threshold $\tau_{cs}$. 
For the typical AP located at $\nbx_0 \in \R^2$, the contention region is $B_{d_{\tt inh}}(\nbx_0)$, which is defined as a circular region of radius $d_{\tt inh}$ centered at the AP.  
We define this distance such that $\tau_{cs} = P_t  l(d_{\tt inh})$.
In Fig.~\ref{fig:system_illus}, we depict the contention region for the typical AP using a dotted black circle of radius $d_{\tt inh}$.
Further, we present the service region of all the active APs through red circles of radius $r_{\tt inh} = d_{\tt inh}/2$. Note that each active AP can be associated with a non-overlapping circular service region in $\R^2$ of size $\pi r_{\tt inh}^2$. 

{\bf \em C. Point process of the active APs:}
In this work, we analyze the system performance for a single representative slot.  
From the description of the back-off process, one can interpret that the APs arrive in the system at random times within a slot based on their back-off times. Further, the arriving locations are random in $\R^2$ based on the underlying PPP $\Phi_A$. An AP is allowed to access the medium if at its time of arrival there are no APs in its contention domain. 
Let us denote the set of active APs at time $t_0 (< t_b)$ by $\Psi_{t_0}$, which is defined as
\begin{align*}
\Psi_{t_0}  = \{\nbx_i: \nbx_i \in \Phi_A, t_i < t_0, |\Psi_{t_i} \cap B_{d_{\tt inh}}(\nbx_i)| = 0 \},
\end{align*}
where the last condition captures that for the AP at $\nbx_i$ with a back-off time $t_i$ to transmit, there should be no active APs in its circular contention region $B_{d_{\tt inh}}(\nbx_i)$.
{\em The point process that  exactly models $\Psi_{t_0}$ is the RSA process~\cite{busson2009,nguyen2012} as will be clear from its definition presented in the next section.}

Now, let us denote $\ncalI_0$ as the medium access indicator of the typical AP with a back-off time $t_0$. If the AP gets to transmit in the slot then $\ncalI_0 = 1$,  otherwise $\ncalI_0 = 0$.
We define 
{\small \begin{align*}
\ncalI_0 = \prod_{\nbx_j \in \Psi_{t_0}} \mathbf{1}\left(d_{0j} \geq d_{\tt inh}\right),
\end{align*}}
where $d_{ij} = \|\nbx_i - \nbx_j\|$.
Using the above definition, at the end of the back-off period, the set of all active APs is 
\begin{align*}
\Psi_{t_b} = \Psi = \{\nbx_i \in \Phi_A : \ncalI_i = 1\}. \numberthis
\label{eq:Psi_T}
\end{align*}
For the rest of the paper, we consider $t_b = 1$.

{\bf \em D. Performance metrics:}
Conditioned on the event that the typical AP at $\nbx_0$ gets to transmit, the DL $\sinr$ for its serving user at $\nbu_0$ is given as 
\begin{align*}
\sinr_0 = \frac{P_t h(\nbu_0, \nbx_0) l(\|\nbx_0 - \nbu_0\|)}{\sum_{\nbx_j \in \Psi^0} P_t h(\nbx_j, \nbu_0) l(\|\nbx_j - \nbu_0\|) + \sigma^2}, \numberthis
\label{eq:sinr_typ_link}
\end{align*}
where $\Psi^0 = \Psi \setminus \{\nbx_0\}$ and $\sigma^2$ is the noise power over the system bandwidth.
Let us denote the random distance $\|\nbx_0 - \nbu_0\| = R_0$. Based on our earlier assumption, the ${\tt PDF}$ of $R_0$ is 
\begin{align*}
f_{R_0}(r_0) = {2 r_0}/{r_{\tt inh}^2}, \quad 0 < r_0 \leq r_{\tt inh}, \numberthis
\label{eq:ServDist}
\end{align*}
where $r_{\tt inh} = d_{\tt inh}/2$. Note that the coverage probability analysis is valid for any other distance distribution as long as the closed-access system assumption in maintained.

In this work, we limit our focus to the following two metrics. It is straightforward to extend the derived theoretical results for other metrics such as average spatial throughput.  

{\em 1) Medium access probability:} The ${\tt MAP}$ of the typical AP at $\nbx_0 \in \Phi_A$ is defined as 
\begin{align*}
\ncalM_0 = \nbbP[\ncalI_0 = 1] = \nbbE\bigg[\prod_{\nbx_j \in \Psi_{t_0}} \mathbf{1}\left(d_{0j} \geq d_{\tt inh}\right)\bigg]. \numberthis
\label{eq:MapDefn}
\end{align*}

{\em 2) $\sinr$ coverage probability:} As explained in Sec.~\ref{sec:SysMod}.B, if the typical AP is active, then the data transmission window is a random variable that is uniformly distributed in $[t_d, t_d + t_b)$. Assuming the back-off time of the typical AP to be $t_0$, the number of active interfering APs keeps increasing during $(t_0, t_b]$ and does not change during $(t_b, t_b + t_d]$. Since the latter window corresponds to the worst-case interference scenario, we define the $\sinr$ coverage probability of the typical link for this window. For a threshold $\beta$, it is given as 
\begin{align*}
\mathrm{P}_c(\beta)  = \nbbP\bigg[ \frac{P_t h_0 l(r_0)}{\sum_{\nbx_i \in \Psi^0} P_t h_i l(r_i) + \sigma^2} \geq \beta \bigg], \numberthis
\label{eq:SinrDefn}
\end{align*}
where  $r_i = \|\nbx_i - \nbu_0\|$, $h_i = h(\nbx_i, \nbu_0)$, and $\Psi^0 = \Psi \setminus \{\nbx_0\}$.

\begin{remark}\label{rem:PPPaprx}
To analytically quantify these metrics, we need to use the statistical properties of $\Psi_{t_0}$, which is an RSA process.
Since the exact characterization of the process is intractable, we focus on deriving approximate results exploiting the first and second-order statistics, namely density and ${\tt PCF}$ of the process. The density result helps derive the ${\tt MAP}$. Further, to obtain the coverage probability result, we approximate $\Psi^0$ as a non-homogeneous PPP due to its unparalleled tractability. This requires information regarding the density as well as the ${\tt PCF}$ of the RSA process that are derived next.
\end{remark}

\vspace{-0.2cm}

\section{A brief overview of the RSA process}
In the two-dimensional (2D) space, the RSA process is defined as a stochastic space-time process, where circles of a certain radius sequentially arrive at random locations in $\R^2$ such that any arriving circle cannot overlap with already existing circles. 
To be specific, let $\Phi$ be a homogeneous space-time PPP on $\R^2 \times [0, 1]$ that contains the centers of circles of radius $r_{\tt inh}$. These circles arrive at the rate of $\lambda_{\Phi}$ per unit area. 
At an arbitrary time $ t \in (0, 1)$, a circle arriving at $\nbx \in \R^2$ is retained if there are no other circle centers in its inhibition region ${B}_{d_{\tt inh}}(\nbx)$. 
Let $\Phi(t)$ be the point process that contains all the centers (both retained and discarded) of $\Phi$ at an arbitrary time $0 < t \leq 1$.
From the definition of $\Phi$, $\Phi(t)$ is a homogeneous PPP with density $\lambda_{\Phi}t$.
Let $\Psi(t) (\subseteq \Phi(t))$ be the corresponding RSA process that has all the {\em retained} points and its density be $\rho(t)$.
If $\lambda_{\Phi}$ is high enough, then beyond a certain time there will be no more empty space left to accommodate a new circle.
In the RSA literature, this is known as the {\em jamming limit}, where the fraction of area covered by the circles is $\rho(1)  \pi r_{\tt inh}^2 \approx 0.547$ as  $\lambda_{\Phi} \rightarrow \infty$ and the corresponding density $\rho(1)$ is the {\em jamming} density~\cite{talbot2000car}.

{\bf \em A. Density of the RSA process:}
Due to the {\em infinite memory} of the RSA process, characterizing its density is challenging.
Nevertheless, many works in the statistical physics literature provide accurate approximation results~\cite{schaaf1989,schaaf1989kinetics, dickman1991}. 
In the following lemma, we present one such result for the density estimation of $\Psi(t)$.

\begin{lemma}\label{lem:RSA_Density}
The density $\rho(t)$ of the point process $\Psi(t)$ is obtained by solving the following equation~\cite{schaaf1989} with the initial condition $\rho(0) = 0$:
\begin{align*}
\int \frac{{\rm d}\rho(t)}{\phi(\kappa \rho(t))} = \frac{\lambda_{\Phi}}{\kappa} t + C, \numberthis
\label{eq:DiffEqun}
\end{align*}
where $\kappa =  \pi r_{\tt inh}^2$ is the unique area covered by a circle, $\kappa \rho(t)$ is the fraction of the area that is covered by the retained circles at time $t$, $\phi(\kappa \rho(t))$ is the probability that a circle arriving at an arbitrary location in $\R^2$ is retained at time $t$, and $C$ is the integration constant.
The series expansion of the retention probability in terms of the density $\rho(t)$ is given as~\cite[Eq.~30]{schaaf1989}
{\small \begin{align*}
\phi(\kappa \rho(t)) =  & 1 - 4 \pi d_{\tt inh}^2 \rho(t) + \frac{\rho(t)^2}{2} \int\limits_{d_{\tt inh}}^{2 d_{\tt inh}} 4 \pi r A_2(r) {\rm d}r \\
& + \frac{\rho(t)^3}{3} \int\limits_{d_{\tt inh}}^{2 d_{\tt inh}} 2 \pi r A_2^2(r) {\rm d}r - S_3^{\tt eq} + O(\rho(t)^4), \numberthis
\label{eq:ExclusionProb}
\end{align*}}
\noindent where $S_3^{\tt eq} = \frac{\rho(t)^3}{8} \pi \left(\sqrt{3} \pi - \frac{14}{3}\right)d_{\tt inh}^6$, $A_2(r)$ is the area of intersection of two circles of radius $d_{\tt inh}$ whose centers are separated by distance $r$.
\end{lemma}
\begin{IEEEproof}
For the proof, please refer to~\cite{schaaf1989}. 
\end{IEEEproof}


While the above result is accurate until intermediate coverage (35\%-40\% occupied area by the admitted circles), the results for the jamming limit are well-known thanks to the existing simulation-based studies~\cite{hinrichsen1986}. Using the above Lemma along with the knowledge of jamming limit coverage, authors in~\cite{schaaf1989} present a fitting function for the fraction of covered area that is fairly accurate for the entire coverage range. This unified equation is given as  $\phi_{\tt FIT}(\kappa \rho(t)) =$
\begin{align*}
 (1 + b_1 x(t) + b_2 x(t)^2 + b_3 x(t)^3)(1-x(t)^3), \numberthis
\label{eq:FitFun}
\end{align*}
where $x(t) = \kappa \rho(t)/0.5474$. The coefficients $b_1, b_2$ and $b_3$ are obtained by matching the order of $\rho(t)$ in \eqref{eq:ExclusionProb} and \eqref{eq:FitFun}.
The density at time $0 < t \leq 1$ can be obtained by solving \eqref{eq:DiffEqun}.

{\bf \em B. {\tt PCF} of the RSA Process:}
Informally, the ${\tt PCF}$ denoted as $g_2(\nbr_1, \nbr_2; t)$ describes the likelihood of finding a point of the process at $\nbr_2$ given that there is a point at $\nbr_1$.
Since RSA is a motion-invariant process, the ${\tt PCF}$ is a function of distance between two locations and independent of absolute locations~\cite{Haenggi2013}. 
Hence, we define $g_2(\nbr_1, \nbr_2; t) \coloneqq g_2(|\nbr_2 - \nbr_1|;t) = g(r;t)$.
Similar to the derivation of the density result, approximated ${\tt PCF}$ result can be obtained using the differential equation framework. 
Since RSA is a stationary process, the second-moment density or the two-particle density is given as $\rho_2(\nbr_1, \nbr_2; t) = \rho(t)^2 g_2(\nbr_1, \nbr_2; t) $. By definition, $\rho_2(\nbr_1, \nbr_2; t) {\rm d}\nbr_1 {\rm d}\nbr_2$ is equal to the probability that the center of one unspecified point can be found in ${\rm d}\nbr_1$ at $\nbr_1$ and another unspecified point in ${\rm d}\nbr_2$ at $\nbr_2$.
Creation of a new pair of particles in an infinitesimally small time window $(t, t + {\rm d}t]$ can occur either by (1) addition of a circle at $\nbr_2$ conditioned on already existing circle at $\nbr_1$, or (2) vice-versa. Hence, the rate of increase of the two-particle density is
\begin{align*}
\frac{\partial \rho(t)^2 g_2(\nbr_1, \nbr_2; t)}{\partial t} = \lambda_{\Phi} \left[\phi(\nbr_1, \nbr_2^0; t) + \phi(\nbr_2, \nbr_1^0; t)\right], \numberthis
\label{eq:PCF_RateChange}
\end{align*}
where $\phi(\nbr_i, \nbr_j^0; t)$ is the probability of finding a circle centered at $\nbr_i$ and empty space at $\nbr_j$ for addition of another circle.
Again due to motion-invariance, $\phi(\nbr_i, \nbr_j^0; t)$ is only a function of the relative distance $|\nbr_i - \nbr_j|$.
Based on already existing results~\cite{reiss1959, tarjus1991}, one can express $\phi(\nbr_i, \nbr_j^0; t) = $
\begin{align*}
& (1 + f_{ij}) \phi(\kappa \rho(t)) \bigg[ \rho(t) + \sum_{s = 1}^{\infty} \frac{\rho(t)^{s + 1}}{s!} \int \ldots \int f_{jk_1} \ldots f_{jk_s} \\ 
& g_{s + 1}(\nbr_i, \nbr_{k_1}, \ldots \nbr_{k_s}; t) {\rm d}\nbr_{k_1} \ldots {\rm d}\nbr_{k_s}\bigg], \numberthis
\label{eq:Condphi}
\end{align*}
where $f_{ij} = f(|\nbr_i - \nbr_j|) = -1$ if $|\nbr_i - \nbr_j| < d_{\tt inh}$ and $f_{ij} = 0$ if   $|\nbr_i - \nbr_j| \geq d_{\tt inh}$, $g_n(\cdot)$ is the $n$-particle correlation function. In the 2D space, the sum over $s$ can be reduced to six terms as a circle can have maximum six neighboring circles.
The ${\tt PCF}$ can be exactly characterized by substituting \eqref{eq:Condphi} in \eqref{eq:PCF_RateChange}. However, as evident from \eqref{eq:Condphi},  without the knowledge of the $n$-particle correlation function for $n \leq 7$, this is not possible.
Therefore, in~\cite{boyer1995}, a numerical approach based on {\em first order density approximation} of $\phi(\nbr_i, \nbr_j^0; t)$ is presented to estimate the ${\tt PCF}$ that is outlined in the Appendix. For a better understanding of the technical arguments presented int the Appendix, the readers may refer to~\cite{tarjus1991}.

\subsubsection{Exponential regression-based curve-fitting}
For tractable analysis, it is convenient to approximate the ${\tt PCF}$ in closed-form, as has been commonly done in the stochastic geometry literature, e.g., for the cellular uplink analysis~\cite{haenggi2017user}.  
We use the following parametric function for this: 
\begin{align*}
g(r; \rho(t)) = 1 + c_1 \exp\left(-c_2 \left({r}/{d_{\tt inh}} - 1 \right)\right), r \geq d_{\tt inh}, \numberthis
\label{eq:PCF_Gen}
\end{align*}
where $c_1$ and $c_2$ are functions of RSA density/time that can be obtained by curve-fitting with respect to the numerically obtained ${\tt PCF}$.
{This choice of exponential function is motivated by the super-exponential decay of the ${\tt PCF}$ of the RSA process~\cite{bonnier1994}.}
Since for the RSA process, the ${\tt PCF}$ is scale-invariant, the values for $c_1$ and $c_2$ need to be determined only once for different densities and later can be reused.
Further, note that the numerically obtained ${\tt PCF}$ as outlined in the Appendix is accurate for low to intermediate coverage ($\leq 40\%$) owing to the first order density approximation in \eqref{eq:Y2c2h2}. 
Therefore, to obtain more accurate estimates of  $c_1$ and $c_2$, we curve-fit \eqref{eq:PCF_Gen} with respect to simulated data using exponential regression.
These values are reported in Table~\ref{tab:c1c2_pcf}.

\begin{table*}
\centering
\caption{Values of $c_1$ and $c_2$ used in the ${\tt PCF}$ for different fractions of the total occupied area.}
\begin{tabular}{|c|c|c|c|c|c|c|c|c|c|c|}
\hline
Occupied area
$(\kappa \rho(t))$ & 0.1  & 0.15 & 0.2  & 0.25 & 0.3  & 0.35 & 0.4  & 0.45 & 0.5 & 0.547 \\ \hline
$c_1$       	   & 0.14 & 0.2  & 0.28 & 0.41 & 0.47 & 0.66 & 0.87 & 1.42 &  1.83   &2.5   \\ \hline
$c_2$   		   & 2    & 2.8  & 2.25 & 4    & 3.4  & 3.96 & 4.38 & 5.92 &  6.78   &7.24   \\ \hline
\end{tabular}
\label{tab:c1c2_pcf}
\end{table*}

{\bf \em C. The RSA process as a non-homogeneous PPP:}
As mentioned in Remark~\ref{rem:PPPaprx}, using the density and ${\tt PCF}$ results, we approximate the RSA process as a non-homogeneous PPP.
\begin{prop}\label{prop:nonPPP}
Conditioned on the location of the typical point of the RSA process $\Psi(t)$, it can be approximated as a non-homogeneous PPP $\tilde{\Psi}^0(t)$ with the following distance-dependent density function:
\begin{align*}
{\lambda}_{\tilde{\Psi}}(r, \rho(t)) = \rho(t) g(r; \rho(t)).
\end{align*}
\end{prop} 
\begin{IEEEproof}
For any function $f: \R^2 \rightarrow R^+$, 
{\small\begin{align*}
& \sum_{\nbx \in \Psi(t)} f(\nbx) = \sum_{\nbx \in \tilde{\Psi}^0(t)} f(\nbx) \Rightarrow \rho(t) \int\limits_{\nbx \in R^2} f(\nbx) g(\|\nbx\| \sqrt{\rho(t)}; \rho(t)) \\ & = \int_{\nbx \in R^2} f(\nbx) {\lambda}_{\tilde{\Psi}}(\|\nbx\|, \rho(t)),
\end{align*}
}

\noindent where the second step follows from the application of Campbell's theorem and replacing the intensity measure by the reduced second factorial moment measure~\cite{Haenggi2013}.
We obtain the final result using the fact that the ${\tt PCF}$ of the RSA process is scale-invariant. Hence, $g(r \sqrt{\rho(t)}; \rho(t)) = g(r; \rho(t))$.
\end{IEEEproof}

\vspace{-0.25cm}
\section{Application to the CSMA Network Analysis}

As mentioned in Sec.~\ref{sec:SysMod}, we have two quantities of interest: (1) the ${\tt MAP}$ of the typical AP in $\Phi_A$ and (2) the $\sinr$ coverage probability conditioned on the fact that the typical AP transmits. 
Next, we begin our discussion with the ${\tt MAP}$. 

{\bf \em A. The ${\tt MAP}$ of the typical AP:}
The ${\tt MAP}$ of the typical AP can be determined using the results from Lemma~\ref{lem:RSA_Density} that is presented in the following proposition.
\begin{prop}\label{prop:MAP}
The ${\tt MAP}$ of the typical AP is given as 
\begin{align*}
\ncalM_0 = \int_0^1 \phi(\kappa \rho(t_0)) {\rm d}t_0,
\end{align*}
where $\phi(\kappa \rho(t_0))$ and $\rho(t_0)$ are obtained using the results in Lemma~\ref{lem:RSA_Density} with $d_{\tt inh} = \left({P_t}/{\tau_{cs}}\right)^{1/\alpha}$.
\end{prop}
\begin{IEEEproof}
Let the typical AP at $\nbx_0$ has a back-off time $t_0$. 
The ${\tt MAP}$ of this AP as defined in \eqref{eq:MapDefn} is
{\small
\begin{align*}
\ncalM_0 = \nbbE[\ncalI_0] = & \nbbE\bigg[\prod_{\nbx_j \in \Psi_{t_0}} \mathbf{1}\left(d_{0j} > d_{\tt inh}\right)\bigg].
\end{align*}
}
\noindent At time $t_0$, the set of active transmitting APs the network is $\Psi_{t_0}$. If there are no active APs in the contention domain of the typical AP, i.e. $|{\Psi}_{t_0} \cap B_{d_{\tt inh}}(\nbx_0)| = 0$, then  the typical AP transmits.
Hence, the probability of the event that the typical AP transmits is equivalent to the probability of finding a empty circular region of diameter $d_{\tt inh}$ in ${\Psi}_{t_0}$. This probability is essentially captured by $\phi(\kappa \rho(t_0))$ given in Lemma~\ref{lem:RSA_Density} and \eqref{eq:FitFun}. 
Since $t_0$ is uniformly distributed over (0, 1], we get the final expression in the proposition by deconditioning over $t_0$. 
\end{IEEEproof}

{\bf \em B. $\sinr$ Coverage probability:}
In the following proposition we present the $\sinr$ coverage probability of the typical link. 
\begin{prop}\label{prop:coverage}
Conditioned on the fact that the typical AP transmits, the $\sinr$ coverage probability of the typical link in the network for a threshold $\beta$ is given as 
\begin{align*}
\mathrm{P}_c(\beta) =  & \int_{r_0 = 0}^{r_{\tt inh}} \exp\left(-\frac{\beta \sigma^2}{P_t l(r_0)}\right)  \\
& \exp\left(- \lambda_{{\Psi}}\int\limits_{r=0}^{\infty} \int\limits_{\theta = 0}^{2 \pi} \frac{\eta(r, r_0,\theta)}{1 + \frac{l(r_0)}{\beta l(r)}}{\rm d}\theta r {\rm d}r  \right) \frac{2 r_0}{r_{\tt inh}^2} {\rm d}r_0,
\end{align*}
where $\eta(r, r_0,  \theta) = g(\sqrt{r^2 + r_0^2 - 2 r r_0 \cos(\theta)})$.
\end{prop}
\begin{IEEEproof}
Using Proposition~\ref{prop:nonPPP}, ${\Psi}^{0}$ can be approximated as a non-homogeneous PPP with density $\lambda_{{\Psi}^0}(r) = \lambda_{{\Psi}}  g(r) =$
\begin{align*}
 \lambda_{{\Psi}} \left(1 + c_1 \exp\left(-c_2 \left({r}/{d_{\tt inh}} - 1 \right)\right)\right), \quad r \geq d_{\tt inh},
\end{align*}
where $\lambda_{{\Psi}} (= \rho(1))$ is determined using Lemma~\ref{lem:RSA_Density}. Further,  $c_1$ and $c_2$ are obtained from Table~\ref{tab:c1c2_pcf} depending on $\lambda_{{\Psi}}$. Now, the coverage probability can be expressed as 
\begin{align*}
\mathrm{P}_c(\beta) & = \nbbP\bigg[ \frac{P_t h_0 l(r_0)}{\sum_{{\nbx}_i \in {\Psi} \setminus \{{\nbx}_0\}} P_t h_i l(\|{\nbx}_i\|) + \sigma^2} \geq \beta \bigg] \\
& = \nbbP\bigg[h_0 \geq \sum_{{\nbx}_i \in {\Psi}^0 } \beta \frac{h_i l(\|{\nbx}_i\|)}{l(r_0)} + \frac{\beta \sigma^2}{P_t l(r_0)}\bigg] \\
& \stackrel{(a)}{=} \nbbE_{r_0}\bigg[e^{- \frac{\beta \sigma^2}{P_t l(r_0)}}\bigg] \nbbE\bigg[ \prod_{{\nbx}_i \in {\Psi}^0} \nbbE_{h_i}\left[e^{- \frac{\beta h_i l(\|\nbx_i\|)}{l(r_0)}}\right]\bigg] \\
& \stackrel{(b)}{=} \nbbE_{r_0}\bigg[e^{- \frac{\beta \sigma^2}{P_t l(r_0)}} \bigg] \nbbE\bigg[ \prod_{{\nbx}_i \in {\Psi}^0} (1 + \beta l(\|\nbx_i\|)/l(r_0))^{-1}\bigg],
\end{align*}
where $(a)$ follows from using the ${\tt CDF}$ of the exponential fading random variable $h_0$, $(b)$ follows from the moment generating function of an exponential random variable, and the final result follows from applying the ${\tt PGFL}$ of a PPP and deconditioning over the serving distance $r_0$ of the typical link.
\end{IEEEproof}

\vspace{-0.25cm}

\section{Results and Conclusion}
In this section, we validate the accuracy of the theoretical results by comparing them with NS-2 simulations and extensive Monte Carlo simulations.  

\begin{figure*}[!htb]
\centering
\begin{subfigure}{0.31\textwidth}
  \centering
  \includegraphics[width=1\linewidth]{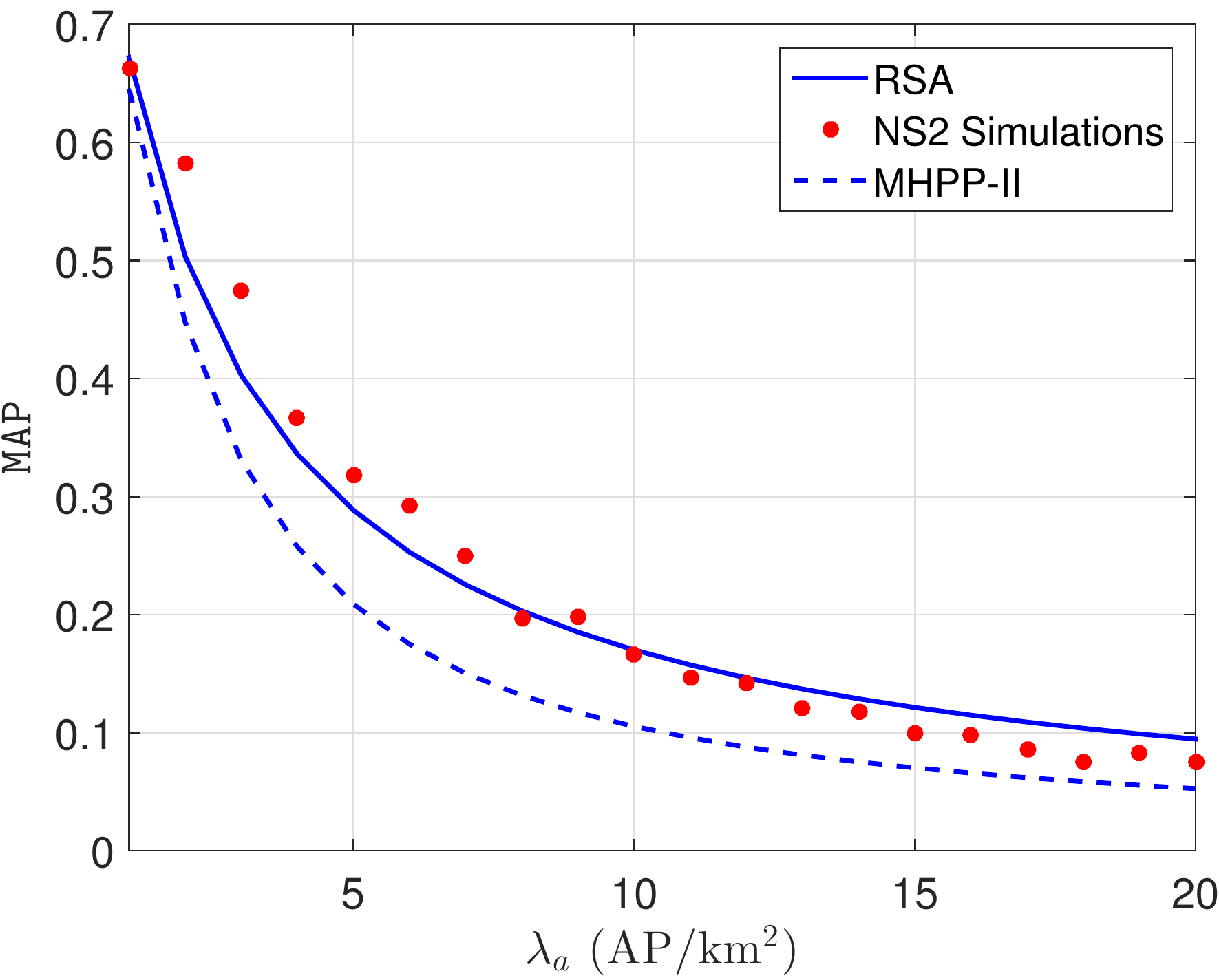}
\end{subfigure}
\begin{subfigure}{0.32\textwidth}
  \centering
  \includegraphics[width=1\linewidth]{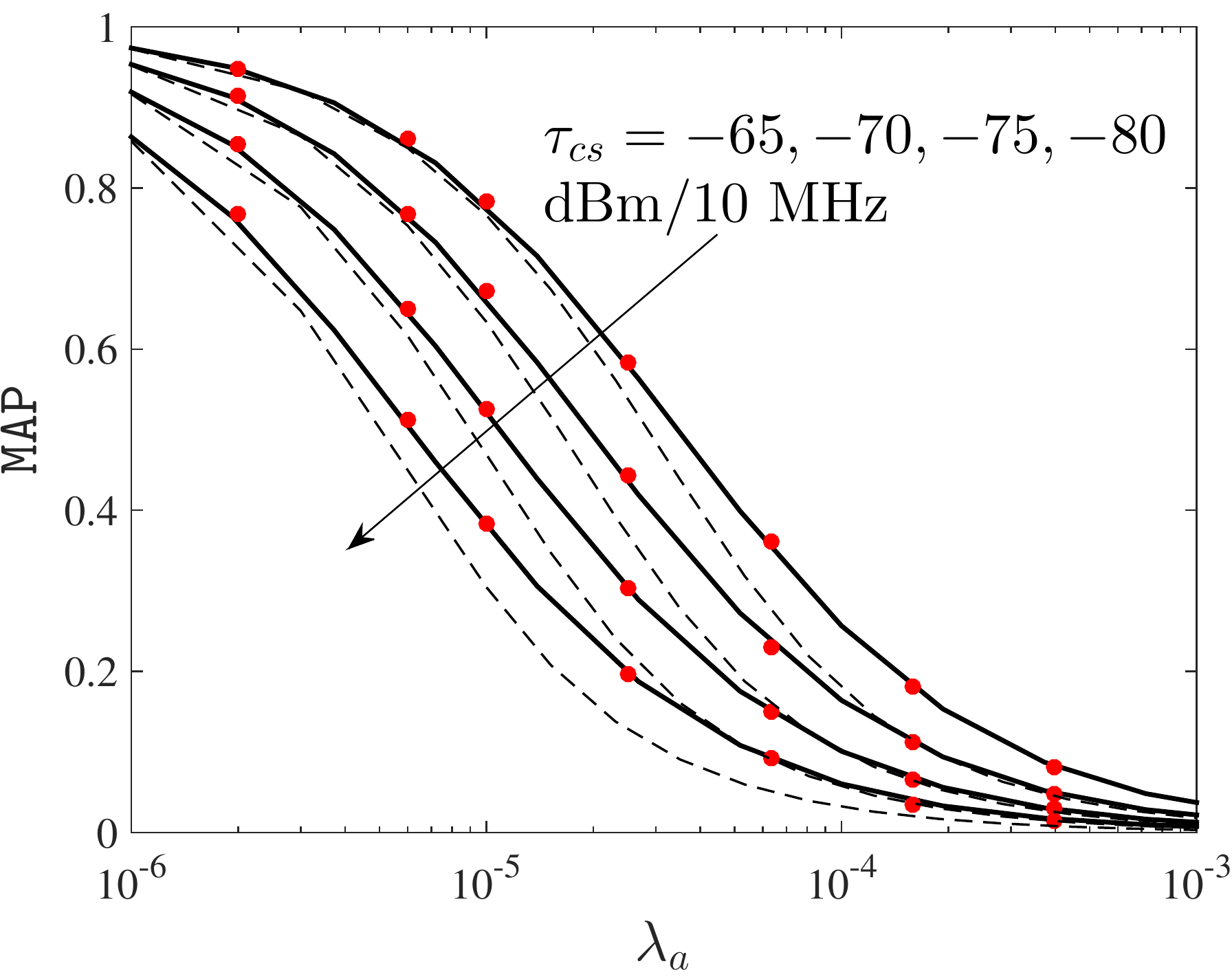}
\end{subfigure}%
\begin{subfigure}{0.30\textwidth}
  \centering
  \includegraphics[width=1\linewidth]{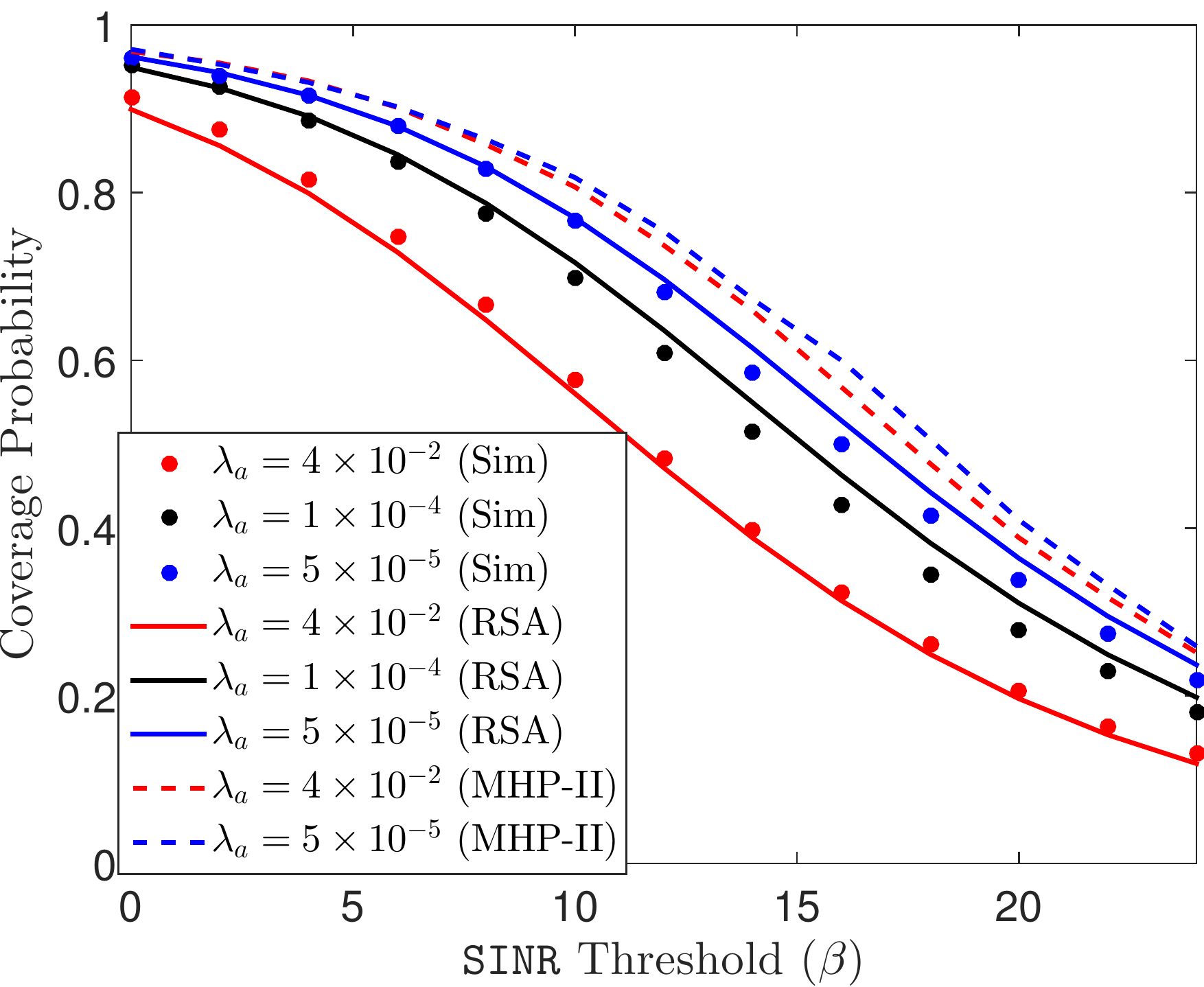}
\end{subfigure}%
\vspace{-0.25cm}
\caption{\footnotesize ({\em Left}) Validation of the accuracy of the RSA-based approach for modeling of CSMA network using NS-2 simulations. ({\em Center}) ${\tt MAP}$ as a function of AP density $\lambda_a$. Markers represent Monte Carlo simulations, solid lines represent theoretical ${\tt MAP}$ of the RSA-based approach using Proposition~\ref{prop:MAP}, and dashed lines represent the ${\tt MAP}$ of the MHPP-II-based approach using \eqref{eq:MAP_mhppII}. ({\em Right}) Coverage probability as a function $\sinr$ threshold for different $\lambda_a$. For the Center and Right figures, $P_t = 20$ dBm/10 MHz, $\alpha = 4$. For the Right figure, $\tau_{cs} = -65 $ dBm/10 MHz.}
\vspace{-0.25cm}
\label{fig:results}
\end{figure*}

{\bf \em A. Validation of the ${\tt MAP}$ result:}
First, we validate the accuracy of modeling a CSMA network with the RSA process by comparing the theoretical ${\tt MAP}$ result with NS-2 simulations.
We use the simulation results reported in~\cite{nguyen2007}. The NS-2 simulation setup that is used to generate the ${\tt MAP}$ result is as follows: the simulation area is considered to be $4 \times 4$ km$^2$. The AP locations are selected randomly and uniformly in the simulation area and each AP is associated with user data protocol (UDP) flows of constant rate 1 Mbps to emulate the saturated traffic condition. Further, all flows start to transmit at the same time. For this simulation, the propagation model is selected to be TwoRayGround and the inhibition distance $d_{\tt inh}$ is 550 m. To obtain the ${\tt MAP}$, the fraction of APs that transmit without collision is obtained during several epochs of a simulation run. 
In Fig.~\ref{fig:results} (left), for the above-mentioned setup, the ${\tt MAP}$ after 50 simulation runs is presented.
To validate the accuracy of Proposition~\ref{prop:MAP} that corresponds to the ${\tt MAP}$ obtained from the RSA-based approach, we compare it with the NS-2 simulations.
Further, to highlight the improved accuracy of the RSA-based approach over MHPP-II-based approach, we also plot the ${\tt MAP}$ of the MHPP-II-based approach that is given as~\cite{nguyen2007} 
\begin{align*}
\ncalM_{\tt 0, M}  = \frac{1 - \exp(- \pi \lambda_a d_{\tt inh}^2)}{\pi \lambda_a d_{\tt inh}^2}. \numberthis
\label{eq:MAP_mhppII}
\end{align*}
From Fig.~\ref{fig:results} (left), we conclude that the RSA-based approach more accurately describes the CSMA network compared to the MHPP-II-based approach.

We further validate the accuracy of the theoretical ${\tt MAP}$ results for denser networks using Monte Carlo simulations. 
Using the stationarity property of the PPP, we consider the typical AP is located at the origin.
For this simulation setup, we consider a circular service region of radius 1500 m centered at the typical AP.
Rest of the AP locations are dropped uniformly at random in this service region. 
Each AP is associated with a back-off time that is uniformly distributed in $(0, 1]$.
Based on its back-off time, if the typical AP gets to transmit, we say that the AP has successfully accessed the medium.
We repeat this process for $10^4$ times to generate the ${\tt MAP}$ result reported in Fig.~\ref{fig:results} (center) for a given density $\lambda_a$. 
As observed from the figure, the theoretical results (from Proposition~\ref{prop:MAP}) and the simulation results are remarkably close.
We also compare the ${\tt MAP}$ result of MHPP-II based modeling using \eqref{eq:MAP_mhppII}.
From Fig.~\ref{fig:results} (left and center), we observe that RSA-based modeling of the CSMA network is more accurate compared to the MHPP-II based modeling used in the literature.

{\bf \em B. Validation of the $\sinr$ coverage probability result:}
In order to validate the coverage probability result, we follow the same  Monte Carlo simulations method as described above. In addition, in each drop, we consider that the location of the user served by the typical AP is uniformly distributed within the circular region $B_{r_{\tt inh}} (\nbo)$. We consider $P_t =  20$ dBm over a 10 MHz system bandwidth and the carrier sense threshold $\tau_{cs}  = -65$ dBm. We run the Monte Carlo simulations $10^4$ times. The DL user $\sinr_0$ is calculated using \eqref{eq:sinr_typ_link} with the condition that the typical AP gets access to the channel.
The simulation-based coverage probability is obtained by taking the ratio of the number of times we get $\sinr_0 \geq \beta$ and the total number drops.
We compare the simulation results with the theoretical coverage probability result obtained using Proposition~\ref{prop:coverage}.
As observed in Fig.~\ref{fig:results} (right), the coverage result is more accurate for a dense network. We also compare the coverage probability result using the MHPP-II-based approach.  We observe that the MHPP-II model significantly overestimates the coverage probability for a denser network. This is not surprising as the MHPP-II process attains the saturation density, which is the maximum density of active transmitter, at a lower value of $\lambda_a$. Hence, it does not represent the actual density of active transmitter in a CSMA network as $\lambda_a$ increases. As a consequence, it significantly underestimates the total network interference.

%

{\bf \em C. Conclusion:}
In conclusion, this paper models a dense WLAN system with CSMA-type medium access protocol using the RSA process.
Leveraging the rich theoretical results from the statistical physics literature, we present approximate analytical results to estimate two key metrics, namely the ${\tt MAP}$ of the typical AP and the $\sinr$ coverage probability of the typical link.
The derived results can be readily extended to analyze more sophisticated metrics such as the average spatial throughput of the system.
Further, these results are also useful in modeling a wireless network that has orthogonal spatial reuse of radio resources, such as frequency reuse in a traditional cellular system.
Moreover, the analysis presented in this work is for a single channel CSMA network. Hence, another interesting direction of future work is to extend the analysis for a multi-channel CSMA network.

\vspace{-0.3cm}

\appendix
\section*{Numerical approximation of the RSA ${\tt PCF}$~\cite{boyer1995}}\label{app:PCF_Numerical}
Let us define a function $Y_2(\nbr_1, \nbr_2^0; \rho(t))$ that represents the conditional probability of finding a point at $\nbr_1$ given there is a circular gap centered at $\nbr_2^0$. Further, there is no constraint on the distance between $\nbr_1$ and $\nbr_2$. Hence, we write
\begin{align*}
\phi(\nbr_1, \nbr_2^0; t)/\phi(\kappa \rho(t)) =  \rho(t)  (1 + f_{12})  Y_2(\nbr_1, \nbr_2^0; \rho(t)), \numberthis
\end{align*}
where 
\begin{align*}
Y_2(\nbr_1, \nbr_2^0; \rho(t)) = & 1 + \sum_{s=1}^\infty \frac{\rho(t)^{s}}{s!} \int \ldots \int f_{2k_1} \ldots f_{2k_s} \\
& g_{s + 1}(\nbr_i, \nbr_{k_1}, \ldots, \nbr_{k_s}; t) {\rm d}\nbr_{k_1} \ldots {\rm d}\nbr_{k_s}.
\end{align*}
A first order approximation for $Y_2(\nbr_1, \nbr_2^0; \rho)$ is~\cite{Given1993} 
{\small \begin{align*}
Y_2(\nbr_1, \nbr_2^0; \rho) & = Y_2(r_{12}, \rho)  = 1 + \rho \int_{\nbr_3} C(r_{13}, \rho) h(r_{32}, \rho) {\rm d}\nbr_3  \\
& = 1 + \rho \int\limits_{\nbr_3} C(|\nbr_3 - \nbr_1|, \rho) h(|\nbr_2 - \nbr_3|, \rho) {\rm d}\nbr_3, \numberthis
\label{eq:Y2c2h2}
\end{align*}
}

\noindent where $C(r_{12}, \rho)$ is the mixed direct correlation function at system density $\rho$ and $h(r_{12}, \rho) = g(r_{12}, \rho)-1$ is the (generic) total pair correlation function. 
With the above approximation, instead of solving \eqref{eq:PCF_RateChange} with \eqref{eq:Condphi}, we can solve \eqref{eq:PCF_RateChange} along with \eqref{eq:Y2c2h2}.
However, since there are three unknown functions, the following additional equation is needed for unique solution
\begin{align*}
C(r_{12}, \rho) = f_{12} Y_2(r_{12}, \rho).
\numberthis
\label{eq:c2Y2}
\end{align*} 
This relationship directly follows from the definition of $C(r_{12}, \rho)$ and $Y_2(r_{12}, \rho)$.
Using \eqref{eq:c2Y2} in \eqref{eq:PCF_RateChange}, we get
{\small \begin{align*}
\frac{1}{2 \rho} \frac{\partial \rho^2 h(r_{12}, \rho)}{\partial \rho} = C(r_{12}, \rho) + \rho \int_{\nbr_3} C(r_{13}, \rho) h(r_{32}, \rho) {\rm d}\nbr_3. \numberthis
\label{eq:h2DiffEq}
\end{align*}
}
\noindent Simultaneously solving \eqref{eq:Y2c2h2}, \eqref{eq:c2Y2}, and \eqref{eq:h2DiffEq}, we get the ${\tt PCF}$.


\vspace{-0.2cm}

\bibliographystyle{IEEEtran}
{\tiny
\bibliography{RSA_CSMA_Lit}}
\end{document}